\documentclass[letterpaper, 12pt]{article}
\usepackage[margin=1in]{geometry}

\usepackage{ltablex}
\usepackage[table, dvipsnames]{xcolor}
\usepackage{float} 
\usepackage{graphicx} 
\usepackage{placeins} 
\usepackage{booktabs} 
\heavyrulewidth{0.25ex} 
\usepackage{array} 
\usepackage{placeins}
\usepackage{tikz}
\usepackage{subcaption}

\usepackage[margin = 1.5cm]{caption}
\usepackage{enumitem}
\setlist{nolistsep}
\setlist[enumerate]{topsep=5pt}
\setlist[itemize]{topsep=5pt}

\usepackage{setspace}
\usepackage{csquotes}
\usepackage{indentfirst} 
\usepackage{amssymb,amsmath,amsthm,enumitem} 
\usepackage[super]{nth}

\usepackage{titlesec}

\usepackage[sort]{natbib}
\renewcommand*{\cite}{\citep}
\usepackage{xcolor}
\usepackage{hyperref}
\hypersetup{
    colorlinks=true,
    linkcolor = blue,
    citecolor = blue,
    urlcolor =blue
}

\title{Conceptualizing educational opportunity hoarding: \\ 
the emergence of hoarding without hoarders}

\author{Joao M. Souto-Maior\footnote{\textbf{Acknowledgments.} This paper resulted from my doctoral dissertation for the Sociology of Education program at  New  York  University.  I  thank  R.  L’Heureux Lewis-McCoy,  Ravi  Shroff,  Samuel  Lucas,  Erez Hatna,  Lisa Stulberg, Marc  Scott, John Skvoretz and Yasmiyn Irizarry  for  helpful  comments  and  feedback.  Earlier  versions  of  this paper  were  presented  at  the 2023 Annual Conference of the International Network of Analytical Sociologists; the 2023 Annual Group Processes Conference; and the 2023 American Sociological Association: Sociology of Education Roundtables. I am thankful to the valuable feedback received at these meetings.}}
\date{%
    Postdoctoral Scholar, Stanford University \\
    \href{joaosm@stanford.edu}{joaosm@stanford.edu} \\ 365 Lasuen St, Stanford, CA 94305\\[3ex]%
    \today
}

\begin{document}
\maketitle

\begin{abstract}
\noindent Social scientists increasingly use the concept of opportunity hoarding to explain the formation of Black-White educational inequalities. However, this concept is often loosely defined, leading to varied interpretations of the inequality-producing mechanisms it captures. To bring clarity to this valuable sociological concept, this theoretical paper, informed by the concept’s original definition and existing empirical research, proposes a more precise definition of opportunity hoarding and formalizes it through a computational model. For concreteness, the model focuses on one context: how White families can hoard access to advanced high school coursework from Black students attending the same school. Through simulations, the paper highlights the necessary and sufficient conditions under which the hoarding of advanced course-taking opportunities emerges. Results demonstrate that, in contrast to traditional accounts, White actors do not need to engage in exclusionary behaviors to hoard valuable resources. Rather, through the byproduct of network segregation and class inequalities, opportunity hoarding can emerge even when individuals act in race-neutral ways --- a process I conceptualize as \textit{hoarding without hoarders}.  
\end{abstract}

\section{Introduction}

A recent body of sociological research emphasizes the role of opportunity hoarding in shaping Black-White educational inequalities \cite[see][for recent literature reviews]{diamond2022opportunity, posey2025advantaged}. This literature demonstrates how this well-known sociological concept can be applied to explain the greater availability of resources in majority-White districts, neighborhoods, and schools \cite{faber2021contemporary, anderson2010imperative, murray2019civil, posey2014middle, rury2018race, rury2011suburban, sattin2020opportunity, kebede2021ethnoracial, goetz2019}, as well as to capture how White families secure better educational opportunities within schools \cite{lewis2014inequality, lewis2015, murray2019civil, murray2020we}.

As originally proposed by \citeauthor{tilly1998durable} (\citeyear[p. 10]{tilly1998durable}), the concept of opportunity hoarding captures processes through which \enquote{members of a categorically bounded network acquire access to a resource that is valuable, renewable, subject to monopoly, supportive of network activities, and enhanced by the network’s modus operandi.} In other words, opportunity hoarding is an umbrella term that captures different micro-level processes through which members of a well-defined social category, such as a racial group, gain (or maintain) access to valuable resources, thereby restricting access for other groups. This broad definition of opportunity hoarding has inspired a wide range of interpretations of the different inequality-producing (or -reproducing) mechanisms that constitute opportunity hoarding. In analyses of Black-White educational disparities (the focus of this paper), two main interpretations stand out:

First, a series of studies, largely rooted in a quantitative tradition, define opportunity hoarding in terms of an observed macro-level outcome: group disparities in access to valuable resources; suggesting that \textit{any} process that produces (or reproduces) Black-White disparities can constitute a form of opportunity hoarding. I refer to this as the \textit{group-disparity} interpretation of opportunity hoarding. For example, scholars articulate that school racial segregation is a form of opportunity hoarding because it leads Black students to attend, on average, lower-quality schools than White students \cite{hanselman2017school, gruijters2024opportunity, reece2016legacy, kebede2021ethnoracial, anderson2010imperative, green2017communities, goetz2019, singer2022race, weathers2022separate}. This view is illustrated by \citeauthor{hanselman2017school}'s (\citeyear{hanselman2017school}) argument that school opportunity hoarding consists of \enquote{the notion that segregation consigns minority students to lower-quality school experiences.} Similarly, within-school tracking is also seen as a form of opportunity hoarding as it often results in the under-representation of Black students in higher-level courses \cite{hanselman2019access, beattie2017tracking, price2021college, perna2015unequal}. As \citeauthor{perna2015unequal} (2015, p. 419) argues, the disproportionate representation of racial minorities in advanced courses \enquote{is a classic example of the way that social inequality can be reproduced through 'opportunity hoarding' by those with privilege.}

Second, and perhaps more commonly, several studies, largely rooted in qualitative and historical traditions, interpret opportunity hoarding in terms of well-defined micro-level processes: the ways in which White individuals (intentionally or unintentionally) engage in race-based exclusionary behaviors, keeping valuable resources within their networks and excluding other groups from access to these resources \cite[e.g.,][]{alvord2021minority, castro2022all, sattin2020opportunity, rury2020creating, lewis2015, lewis2014inequality}. I refer to this as the \textit{exclusionary-behaviors} interpretation of opportunity hoarding.

One kind of exclusionary behavior is defined by what I call \textit{administrative pressures} --- where White families (intentionally or unintentionally) leverage their economic and political influence to allocate resources in ways that benefit White communities, allowing educational systems to operate as \textit{White spaces} \cite{diamond2022opportunity, anderson2022black}. Numerous studies illustrate how organized advocacy by White communities contributes to between-school inequalities by creating or reinforcing geographic, district, or attendance boundaries \cite{castro2022all, grooms2019turbulence, rury2011suburban, rury2018race, rury2020creating}; shaping the structure of school funding systems \cite{alvord2021minority, nations2020racial, steffes2020assessment, posey2016beyond}; and influencing the design of school choice programs \cite{sattin2020opportunity, roda2013school, roda2018school}. Similarly, studies also illustrate how pressures from White families help shape within-school inequalities through organizational practices and routines --- such as biased enforcement of disciplinary practices \cite{diamond2019race, lewis2015} and intensified within-school tracking \cite{lucas2002sociodemographic, joffe2020nice} --- that advantage White students.

Another kind of exclusionary behavior involves the notion of \textit{in-group favoritism} --- the favoring of same-race ties in the sharing of valuable resources, such as information \cite{aboud2003cross}. This in-group favoritism manifests through the prevalence of more meaningful interactions with in-group members, even among individuals already within one’s network \cite{royster2003race, mcdonald2013contributions, ditomaso2013american, currarini2016identity}. To illustrate, note that favors---such as sharing of information about employment opportunities or job recommendations---\enquote{are made for those with whom one identifies, for those who are considered insiders in one’s social networks, and for those whom one believes to be trustworthy enough to return the favor to others in the community without violating the group’s norms.} \cite[p. 69]{ditomaso2013american}. Further, it manifests through the (overt or covert) exclusion of racial minorities from spaces --- such as parent-teacher organizations --- that are central in shaping communities' educational landscapes \cite{posey2017race, lewis2014inequality}.

This theoretical paper argues that these two interpretations --- group-disparity and exclusionary-behaviors --- may fail to capture the full conceptual scope of opportunity hoarding. The group-disparity interpretation is too inclusive, offering limited insight into the specific inequality-producing processes that constitute opportunity hoarding. Conversely, the exclusionary-behaviors interpretation is too restrictive, overlooking how opportunity hoarding can occur even when individuals act in truly race-neutral ways.

I develop this argument in two steps. First, I articulate the proposed interpretation of opportunity hoarding, providing intuition for why these common interpretations face limitations. Second, to bring conceptual precision to this argument and link the micro-level processes of interest to the macro-level pattern being explained \cite{raub2022rigorous, raub2023methodological, manzo2022agent, hedstrom2005dissecting, hedstrom2010causal}, I construct a model of educational opportunity hoarding that details the interpretation proposed here and identifies the necessary and sufficient conditions for opportunity hoarding to emerge.

\section{Defining opportunity hoarding}

Opportunity hoarding is, by definition, an explanation for group-level disparities. An explanation should involve two components: the \textit{explanandum} (the pattern to be explained) and the mechanisms which generate this pattern \cite{elster2006explaining, hedstrom2010causal, hedstrom2009analytical, epstein1996growing, hedstrom1998social, manzo2024antecedents}. In the social sciences, the primary goal is to explain social phenomena. The \textit{explanandum} of interest, therefore, is often defined at the macro-level, capturing aggregate patterns such as spatial distributions and group-level inequalities \cite{hedstrom2010causal}. Further, social mechanisms, broadly defined, open the black box which generates the phenomena of interest by detailing the chain of micro-level processes (such as individual actions and interactions) which can (under well-defined conditions) generate the pattern of interest \cite{hedstrom2010causal, hedstrom2009analytical}. Defining opportunity hoarding, therefore, involves three tasks:

\begin{itemize}
    \item[(1)] define the \textit{explanandum};
    \item[(2)] define the mechanisms (i.e., micro-level processes) of interest; and
    \item[(3)] show the conditions under which the micro-level processes of interest generate the pattern we want to explain.\footnote{To understand the importance of this last step, note that individual behavior does not occur in a vacuum; it is context-dependent. A comprehensive explanatory model must go beyond \enquote{rock-bottom} explanations --- which begin from idealized micro-processes --- and detail the macro-level conditions under which the micro-level mechanisms produce the macro-level outcome \cite{hedstrom2009analytical, coleman1986social}.}
\end{itemize}

To define the \textit{explanandum} (task (1)), consider the context in which the concept of opportunity hoarding was proposed. As \citeauthor{posey2025advantaged} (2025) highlights, one of Tilly's central goals for introducing this concept was to complement dominant explanations for group-level disparities at the time of his writing. In particular, he sought to challenge the individualistic paradigm which emphasized the role of individual attributes --- such as \enquote{human capital, ambition, educational credentials} \cite[p. 22]{tilly1998durable} --- in explaining group disparities. In fact, his central thesis was that \enquote{[l]arge, significant inequalities in advantages among human beings correspond mainly to categorical differences such as black/white, male/female, citizen/foreigner, or Muslim/Jew rather than to individual differences in attributes, propensities or performances} \cite[p. 7]{tilly1998durable}. The concept of opportunity hoarding, therefore, targets a specific \textit{explanandum}: group disparities in access to a valuable resource \textit{that cannot be explained by differences in individual attributes}, such as social class or educational background. For short, we can refer to this \textit{explanandum} as a \textit{group-based penalty}; i.e., the extent to which a lower likelihood of resource access stems from group membership itself, rather than from individual differences.

Here, one clarification is useful. As defined by \citet{tilly1998durable}, opportunity hoarding can refer to both the \textit{emergence or reproduction} of such disparities. That is, it can capture how a group-based penalty is formed or how it is maintained over time. These are clearly different \textit{explananda} and, therefore, it is useful to differentiate between them. In this paper, I am concerned with opportunity hoarding as an explanation for the \textit{emergence (rather than reproduction)} of race-based penalties in access to resources. Future investigations might consider a similar exercise when emphasizing the \textit{reproduction} of race-based penalties over time. 

To define the micro-level processes of interest (task (2)), note that the concept of opportunity hoarding was introduced to emphasize the role of social interactions in producing group-level disparities. As Tilly (\citeyear[p. 236]{tilly1998durable}) writes, the concept highlights that \enquote{[b]onds, not essences, provide the bases of durable inequality.} Following this original definition, opportunity hoarding is concerned with relational (or network-based) micro-level processes --- that is, processes that fundamentally depend on social interactions \cite{emirbayer1997manifesto}.

From these definitions, it follows that racial opportunity hoarding captures \textit{the relational processes that can generate race-based penalties in access to valuable resources.} Further, to fully articulate the concept of opportunity hoarding (task (3)), one must identify the macro-level conditions under which such micro-level processes generate the pattern we seek to explain.

As discussed in the next section, this interpretation also helps clarify the limitations of existing applications of the opportunity hoarding concept.

\subsection{Limitations of existing interpretations}

The group-disparity interpretation faces two problems. First, it proposes an \textit{explanandum} (any group-level disparity) that is too broad. As argued above, opportunity hoarding is specifically concerned with explaining group-level disparity that cannot be attributed to individual-level differences, or, to use Tilly's terminology quoted above, \enquote{essences}. Second, this interpretation is unspecific about the kinds of micro-level processes that opportunity hoarding is concerned with. Because it is so expansive (both in terms of the \textit{explanandum} and of the mechanisms of interest), this interpretation leads the concept of opportunity hoarding to have little meaning, allowing any process that produces any kind of group-level disparity to be defined as opportunity hoarding.

The exclusionary-behaviors interpretation is more specific about the micro-level processes of interest, defining opportunity hoarding in terms of the exclusionary behaviors of White actors --- summarized by the administrative influence and in-group favoritism mechanisms outlined above. However, this interpretation is too narrow, as administrative influence and in-group favoritism are not the only relational processes that can generate race-based penalties in access to valuable educational resources. It overlooks how another relational mechanism --- \textit{the diffusion of network-based resources}, where network-based resources are those acquired through social interactions --- can also produce race-based penalties, even when behaviors are race-neutral, a dynamic I conceptualize as \textit{hoarding without hoarders}.

To unpack this argument, I build on the notion that the concentration of disadvantage within segregated networks can reproduce or exacerbate disparities over time \cite{massey1990american, massey1994migration}, and show (through the model constructed below) that when access to a valuable educational resource (e.g., advanced coursework) depends on access to a network-based resource (e.g., information), the interplay between racially segregated networks and initial group-level differences in access to network-based resources can generate disparities between otherwise comparable individuals of different races, even when actors behave in race-neutral ways.

\subsection{The usefulness of formalization in conceptualizing opportunity hoarding}

In what follows, I will construct an agent-based model\footnote{I focus on the construction of an agent-based model (rather than a purely mathematical model) because this methodology allows for computational rules to be less constrained by mathematical tractability, allowing for more flexible rules and less deterministic outcomes \cite{epstein2006generative}.} to formalize the proposed interpretation of opportunity hoarding as well as the proposed notion of hoarding without hoarders. This formalization has two goals.

First, to foster conceptual precision. Perhaps part of the reason why opportunity hoarding has yielded multiple interpretations is that it was originally articulated verbally. Translating the definitions above into a model helps crystallize claims \cite{raub2023methodological, ylikoski2009illusion}. Specifically, the model will articulate how exclusionary behaviors --- defined by the administrative influence and in-group favoritism mechanisms --- are not necessary to generate race-based penalties in access to educational resources. It will show how the diffusion of network-based resources, regardless of exclusionary behaviors, can constitute a foundation of opportunity hoarding.

Second, to take on task (3), i.e., to show the conditions under which the micro-level processes of interest can generate racial penalties. This is important not only for theoretical precision but also for practical applications of the theoretical framework: understanding these conditions provides pathways to counter the phenomenon (opportunity hoarding) that do not rely on the challenging task of influencing individual behavior. However, sociologists are well aware that it is often difficult to intuitively grasp how network-based dynamics produce specific macro-level patterns. Formalization can help in making these dynamics more transparent \cite{manzo2022agent, stewart2023race, hedstrom2005dissecting, bruch2015agent, schelling1971dynamic}.

For concreteness, the model will focus on one particular \textit{explanandum}: the emergence of a race-based penalty --- a Black penalty and a White premium---in access to high school coursework between students in the same school. Recall that a race-based penalty captures group disparities which are not explained by individual-level differences. I focus on this context because it exemplifies a setting in which all the micro-level processes outlined above --- administrative influence, in-group favoritism, and diffusion of network-based resources --- are identified in the literature, making it an ideal representative case to analyze their roles in the production of opportunity hoarding.

\section{The context of interest: within-school Black-White disparities in access to advanced high school courses}

Racial minorities remain notably underrepresented in advanced high school courses --- such as advanced mathematics, honors, and Advanced Placement \cite{lucas2020race, xu2021college, riegle2018gender}. A central explanation for these disparities is the unequal availability of advanced curricula across schools \cite{roscigno1999race, iatarola2011determinants}. However, even when we compare Black and White students attending the same school, Black students are less likely than White students to enroll in advanced courses \cite{tyson2011, clotfelter2011, mickelson2001subverting, lucas2007race, kelly2009black, souto2024differences, irizarry2021track}.

Explanations for such within-school disparities often focus on the fact that Black and White students differ across relevant attributes --- such as academic background and social class --- at the time of high school entry. However, evidence suggests that Black-White differences in access to advanced coursework might not be fully explained by such individual-level differences \cite{souto2024differences, irizarry2021track}, particularly in racially diverse schools \cite{lucas2007race, kelly2009black}. In fact, as detailed below, the relational processes of administrative influence, diffusion of network-based resources, and in-group favoritism can help explain why disparities could arise even among comparable individuals. In what follows, I review both the individualistic and relational explanations for such within-school disparities.

\subsection{Individualistic explanations}

Below I highlight five key individual-level attributes which can explain within-school Black-White differences in advanced enrollment. 

\subsubsection*{Academic background} 

Due to inequalities in early educational experiences, Black students tend to enter high school with lower academic preparation than White students \cite{souto2024differences, gamoran1987stratification, lucas2007race, kelly2009black}, often not being \enquote{on track} to take many advanced courses \cite{irizarry2021track}. 

\subsubsection*{Differential influence} 

It is common for parents (and students) to try to influence gatekeepers' decisions, advocating for education customization, such as placement in advanced courses \cite{tyson2011, lewis2015, lewis2014inequality}. Because upper-class families can attract more financial resources to their districts and schools \cite{murray2019civil, owens2018income}, teachers and administrators can often feel pressured to be more responsive to the customization requests these families make \cite{calarco2018, weininger2003translating, roda2013school, cucchiara2013marketing, posey2014middle, calarco2020avoiding, sattin2020opportunity, joffe2020nice}. Then, because White families are often from higher social class background than Black families, differential ability to influence gatekeepers' decisions can explain Black-White differences in advanced enrollment.

\subsubsection*{Academic aspirations} 

Black and White students might enter high school with different educational aspirations, having different interests in taking advanced courses \cite{fryer2010empirical, francis2021separate}. This can occur because the decision to enroll in advanced courses is often tied to whether one has been accustomed to enrollment in advanced coursework \cite{tyson2011} as well as to expectations about feelings of social belonging in advanced courses \cite{tyson2011, oconnor2011being, frank2008social}. A brief note on why such racial differences in academic aspirations arise is important. Scholars have hypothesized that such differences result from intrinsic cultural differences --- the argument is that because of perceived lack of opportunities in the labor market, Black communities have developed cultural values that are oppositional to behaviors focused on academic achievement, such as aiming for high grades or taking advanced courses \cite{fordham1986black, austen2005economic}. However, this explanation lacks empirical support, as there is little evidence that students' educational aspirations are exogenously defined \cite{harris2011, tyson2005black, tyson2011}. Rather, more recent theories emphasize that contextual conditions (rather than intrinsic racial differences) underlie the formation of racial differences in educational aspirations \cite{francis2020isolation, francis2021separate, tyson2011, souto2025school}.

\subsubsection*{Access to information} 

Various information resources matter for one's chances of advanced enrollment. Students, for instance, need knowledge about the academic significance of advanced courses; about the steps required to become academically eligible for advanced courses --- such as following the appropriate course sequence, organizing schedules and building positive relationships with teachers in order to secure recommendations \cite{calarco2018, lareau2011, lewis2014inequality, lewis2015, cooper2007structure, crosnoe2001academic}. Parents are central providers of such information and upper-class parents' --- either because of their own prior social and educational experiences or because of their ability to afford private support \cite{mcdonough1997choosing, mcdonough2005counseling, lareau2011} --- are more likely to convey their children with access to such information. Then, since Black black families might often come from a lower social class background, access to information can also explain Black-White differences in advanced enrollment.

\subsubsection*{Social class}

Numerous studies report how social class is a key factor explaining racial differences in course-taking \cite[e.g.,][]{irizarry2021track, rodriguez2019more, kelly2009black, lucas2007race, conger2009explaining, gamoran1987stratification}. The role of social class operates through its correlation with all the attributes identified above. 
Social class is correlated with academic background \cite{reardon2019geography}; differential influence \cite{calarco2018, cucchiara2013marketing, posey2014middle, kelly2004increased}; academic aspirations \cite{tyson2005black}; and access to the information resources \cite{lareau2011, cooper2007structure, bennett2012beyond}. 

\subsection{Relational explanations}

Now, let us consider possible relational explanations --- those that stem from social interactions rather than solely from individual characteristics --- for within-school Black-White differences in advanced enrollment.

\subsubsection*{Administrative pressures and the creation of White spaces}

As noted in the prior section, parents (and students) often to try to influence gatekeepers' decisions and social class can influence one's ability to successfully do so --- providing upper-class students with an individual-level advantage. However, these interactions between parents (or students) and school personnel (teachers, staff, and administrators) do more than confer benefits to individual students; they also shape the institutional environment by reinforcing particular norms and expectations \cite{lewis2014inequality, lewis2015, diamond2019race}. These customization requests, therefore, constitute a form of exclusionary behavior that helps constitute the school environment as a White space \cite{diamond2022opportunity}---an environment where White cultural practices and behaviors are valued and rewarded at higher rates.

To better understand how this unfolds, note that upper-class families are often disproportionately White. When these families engage in self-serving practices and pressure schools to customize educational settings to better serve their preferences, they --- even if unintentionally --- contribute to structuring schools around White norms and values \cite{lewis2014inequality, diamond1999beyond, diamond2022opportunity, diamond2019race, allen2018s}. For instance, as school personnel respond to these demands, they may become more receptive to the cultural styles (e.g., communication patterns, dress, parental involvement strategies) associated with White families \cite{diamond2019race}. This means that when both Black and White families advocate for educational customization --- such as access to advanced coursework---they may face different chances of success \cite{lewis2014inequality}.

\subsubsection*{Diffusion of network-based resources}

Many of the individualistic explanations highlighted above constitute advantages which fully depend on one's own background (e.g., academic background, social class background). However, other factors---academic aspirations and information---can also be acquired through social interactions. With respect to academic aspirations, note that, for many students, the decision to take advanced courses can be a difficult social decision. Moving to advanced courses can involve joining a new set of peers, leading many students to fear that they will not belong socially in this new environment \cite{francis2021separate, tyson2011, oconnor2011being}. Thus, students’ motivation to enroll in advanced coursework can be highly susceptible to the messages they receive from their friends. With respect to information, note that many families cannot rely on their own background or financial resources to gain access to necessary information resources. For these families, access to such information depends on formal or informal interactions with other families in the school community \cite{small2009unanticipated, lewis2014inequality}. The notion of \textit{network-based resources} helps capture these valuable resources, such as academic aspirations and information, which are a function of social interactions. 

\subsubsection*{In-group favoritism}

Importantly, the sharing of such network-based resources is often defined by some level of in-group favoritism, being more likely to occur within members of the same racial group. The sharing of information resources, for instance, tends to be reserved for those with whom one has a strong sense of shared identity, being more likely to occur within same-race ties \cite{ditomaso2013american, lewis2014inequality, mcdonald2013contributions}. Similarly, in the case of academic aspirations, because the constructions of students’ identities in school can often be racialized, students might be more sensitive to the messages received from same-race friends \cite{oconnor2011being, nasir2017stem}. Overall, therefore, in-group favoritism makes the sharing of network-based resources more likely to occur within same-race groups.

\section{The agent-based model}

To articulate the notion of opportunity hoarding in the context of high school course-taking, this agent-based model details how the relational processes outlined in the previous section can generate disparities in advanced enrollment between Black and White students net of the individual-level attributes highlighted above.

\subsection{Overview}

Consider a cohort of 400 agents entering a high school.\footnote{To define an empirically reasonable cohort size, I consider descriptive statistics from New York City high schools \cite{souto2024differences}. Looking at high schools offering AP mathematics courses, the study reports an average \nth{9} enrollment of about 407 students per school (46,890 students over 115 high schools). This is higher than the national average, which, based on the 2017–18 Common Core of Data, is about 200 students. That said, a reasonably large cohort size helps with computational efficiency---i.e., since the number of agents is necessarily discrete, a larger number of agents allows us to more easily vary the percentage of agents with given characteristics across simulations.} Each agent represents a student-parent unit. The model represents a highly stylized high school and, therefore, is defined by several idealizations. First, every agent who enters high school progresses in an ideal grade-promotion trajectory---i.e., the student does not repeat grades or drop out, finishing high school in four years. Second, to concentrate on the dynamics between the two race groups of interest, let this high school be composed only of Black and White students, 50\% each.\footnote{I focus on such a racially heterogeneous high school because these are the kinds of high schools in which the outcome of interest---racial penalties in advanced enrollment---are known to be more salient \cite{lucas2007race, kelly2009black, souto2025school, diette2012whiter}.} Finally, this idealized high school offers only two types of courses: regular and advanced. The advanced course is a limited educational resource, available to only a fraction of students. In the simulations presented here, let the number of available spots in the advanced course be 25\% of the student body (100 spots).\footnote{As done for the number of agents, to define an empirically reasonable value for the size of the advanced course, I consider descriptive results from \cite{souto2024differences}. Looking at one example of advanced high school coursework---AP mathematics courses, defined as AP Calculus AB, AP Calculus BC, and AP Statistics---the study shows that 23.9\% of students who follow a standard grade promotion trajectory enroll in at least one AP mathematics course before finishing high school.}

Informed by the basic structure of Souto-Maior's (\citeyear{souto2025school}) empirically validated model of advanced enrollment, the model captures students' competition for advanced courses using the following logic:

\begin{enumerate}
    \item \textbf{Initialization.} Time step $= 0$ represents the moment at which agents enter high school. At this point, agents are endowed with initial (pre-high school) characteristics and are placed within a network structure, being endowed with undirected network ties with other agents in the school community.
    \item \textbf{Competition for advanced enrollment.} Each time step $> 0$ represents opportunities for interactions between agents and for enrollment in the advanced course.
    \item \textbf{Stopping condition.} Once all available spots in the advanced course are taken, the model stops.
\end{enumerate}

From this basic structure, the model presented here expands on Souto-Maior's (\citeyear{souto2025school}) model to capture all the micro-level processes of interest---administrative influence, in-group favoritism, and diffusion of network-based resources---thus providing a formalization of opportunity hoarding.

\subsection{Variables and parameters}

In what follows, I describe the variables and parameters that define the model. For readability, I use variable names to describe agent-level variables; $\alpha$ parameters to describe the structural conditions that characterize the simulated environment; and $\beta$ parameters to capture the mechanisms of interest.

\subsubsection*{Agent-level variables}

Reflecting the individualistic explanations outlined above, each agent $i$, at the start of the model, is endowed with five characteristics:

\begin{itemize}
\item \textbf{Race}. The race of agent $i$ is captured by dummy variable $\textit{black}_i$, which captures whether the agent is Black. Because there are only Black and White agents in this high school, it follows that if $\textit{black}_i = 0$, the agent is White.
\item \textbf{Academic background}. Continuous variable $\textit{acad-background}_i \in [0, 1]$ captures the level of the agent's academic background.
\item \textbf{Potential to influence course placement decisions}. Continuous variable $\textit{influence-potential}_i$ captures $i$'s potential to influence course placement decisions, i.e., the extent to which the agent can successfully shape gatekeepers' decisions.
\item \textbf{Access to network-based resources (information and academic aspirations)}. Dummy variable $\textit{net-resource}_i$ captures whether agent $i$, at the time of high school entry, has access to the network-based resources that facilitate the navigation of an advanced academic trajectory. For generality, this variable should be interpreted as a composite of all network-based resources (such as information or academic aspirations) that are meaningful to students’ chances of advanced enrollment.
\item \textbf{Social class}. For simplicity, let agents be either from a lower or upper social class. Then, dummy variable $\textit{upper-class}_i$ captures whether agent $i$ is from an upper ($\textit{upper-class}_i = 1$) or lower ($\textit{lower-class}_i = 0$) social class.
\end{itemize}

\subsubsection*{Structural conditions}

Following the empirical reality of most U.S. high schools, Black and White agents---due to current and historical structural inequalities---differ in important ways at the time of high school entry. These disparities manifest both in individual-level attributes and in the structure of agents' social networks.

With respect to individual-level characteristics, the model allows Black and White agents to differ across all the agent-level attributes described above: $\textit{acad-background}_i$, $\textit{influence-potential}_i$, $\textit{net-resource}_i$, and $\textit{upper-class}_i$. Importantly, because the model focuses on the \textit{emergence (rather than the reproduction)} of racial penalties, we seek to simulate racial differences in advanced course-taking among agents which start the model with the same individual-level characteristics. To achieve this, the simulations presented here, informed by the known correlation between social class with all these attributes, consider a stylized context in which all initial Black-White differences are \textit{fully explained by racial differences in social class}. Then, when assessing simulation results, we can simply control for social class to ensure a comparison between agents which are not initially disadvantaged. Overall, this allows us to investigate whether racial penalties can arise \textit{endogenously}, even when race has no direct effect on agents’ initial characteristics.

With respect to social networks, the model incorporates racial homophily, allowing same-race ties to be more likely than different-race ties.

The model is then defined by two structural conditions:

\begin{itemize}
    \item \textbf{Racial differences in social class.} Parameter $\alpha_{\textit{class-ineq}}$ captures the level of Black-White differences in social class background. Let $\alpha_\textit{class-b}$ be the probability that a Black agent is from an upper class and $\alpha_\textit{class-w}$ be the probability that a White agent is from an upper class. Then, let $\alpha_{\textit{class-ineq}} = 1 - \frac{\alpha_\textit{class-b}}{\alpha_\textit{class-w}}$. Consistent with the empirical reality of most U.S. high schools, let $\alpha_\textit{class-w} \geq \alpha_\textit{class-b}$, and so $\alpha_\textit{class-ineq} \in [0,1]$. Intuitively, this means that when $\alpha_\textit{class-ineq} = 0$, there are no Black-White differences in social class background ($\frac{\alpha_\textit{class-b}}{\alpha_\textit{class-w}} = 1$). As $\alpha_\textit{class-ineq}$ increases, Black-White differences in social class background increase (the ratio $\frac{\alpha_\textit{class-b}}{\alpha_\textit{class-w}}$ decreases), reaching its maximum when $\alpha_\textit{class-ineq} = 1$. To initialize other individual attributes at the time of high school entry, based on differences in social class, let: $\textit{acad-background}_i = 0.75 + 0.25 \; \textit{upper-class}_i$,\footnote{This operationalization choice ensures that lower-class students have a nonzero value for $\textit{acad-background}_i$. For upper-class students ($\textit{upper-class}_i = 1$), $\textit{acad-background}_i = 1.0$. For lower-class students ($\textit{upper-class}_i = 0$), $\textit{acad-background}_i = 0.75$.} $\textit{net-resource}_i = \textit{upper-class}_i$, and $\textit{influence-potential}_i = \textit{upper-class}_i$.
    
    \item \textbf{Network racial segregation.} The level of network racial segregation is defined by Coleman's \citeyearpar{coleman1958relational} inbreeding racial homophily index, captured by parameter $\alpha_{\textit{racial-homophily}}$. This index captures the average tendency for same-race ties, net of opportunities for contact. An inbreeding homophily of 0 indicates that agents’ likelihood of forming same-race ties is fully explained by opportunities for contact. For instance, in a school in which Blacks represent 50\% of the student body, inbreeding racial homophily for Black students is 0 if, on average, 50\% of their ties are with same-race peers. \ref{A:net-model} details the computational rules for the formation of agents' networks.
\end{itemize}

\subsubsection*{Micro-level processes of interest} 

The model relies on the following parameters to capture the micro-level processes of interest:

\begin{itemize}
\item \textbf{Administrative pressures.} Let continuous parameter
$\beta_\textit{admin-pressures} \in [0, 1]$ represent the extent to which agents---students and parents---engage in practices to pressure gatekeepers' decisions. If $\beta_\textit{admin-pressures} = 0$, then agents do not display administrative pressure towards gatekeepers. As $\beta_\textit{admin-pressures}$ increases, the level of such administrative pressures increases, reaching its maximum when $\beta_\textit{admin-pressures} = 1$.
\item \textbf{Diffusion of network-based resources.} Let continuous parameter $\beta_\textit{net-resource} \in [0, 1]$ capture the extent to which network-based resources (such as information or students' expectations about social belonging) matter to placement decisions. As will be clear in the description of computational procedures below, when $\beta_\textit{net-resource} = 0$, network-based resources do not matter for advanced enrollment, and thus the diffusion of network-based resources does not influence course-taking dynamics. As $\beta_\textit{net-resource}$ increases, the role of diffusion of network-based resources on course-taking dynamics increases, reaching its maximum when $\beta_\textit{net-resource} = 1$.
\item \textbf{In-group favoritism.} Parameter $\beta_\textit{in-group-favoritism}$ captures the extent to which diffusion of resources varies across same- and different-race ties. Let $\beta_\textit{diffusion-s}$ be the probability of resource diffusion between same-race agents and $\beta_\textit{diffusion-d}$ be the probability of resource diffusion between different-race agents. Then, let $\beta_\textit{in-group-favoritism} = 1 - \frac{\beta_\textit{diffusion-d}}{\beta_\textit{diffusion-s}}$. Consistent with the notion of in-group favoritism, let $\beta_\textit{diffusion-s} \geq \beta_\textit{diffusion-d}$, and so $\beta_\textit{in-group-favoritism} \in [0,1]$. Intuitively, this means that when $\beta_\textit{in-group-favoritism} = 0$, there is no in-group favoritism in social interactions. As $\beta_\textit{in-group-favoritism}$ increases, the salience of in-group favoritism increases, reaching its maximum when $\beta_\textit{in-group-favoritism} = 1$.
\end{itemize}

\subsection{Outcome of interest}

As defined above, opportunity hoarding is an explanation for the emergence of racial penalties in access to a particular resource---i.e., racial disparities in resource access that are unexplained by differences across individual attributes. In the context of this model, the goal is to assess whether (and the extent to which) disparities in advanced course enrollment between Black and White agents emerge even when they enter high school with the same individual characteristics (social class, academic background, access to network-based resources, and potential to influence gatekeepers' decisions).

Because the initialization fully attributes Black-White differences to disparities in social class, controlling for social class allows us to compare individuals who enter high school with the same individual characteristics. Therefore, for the outcome of interest, I concentrate on the simulated differences in advanced enrollment between lower-class Black and White agents once the model stops.

Formally, the outcome of interest is the Black-White relative risk ratio of advanced enrollment among lower-class agents, denoted $R_{bw}$. Let $N_b^l$ and $N_w^l$ represent the number of lower-class Black and White agents in the school, respectively, and let $E_b^l$ and $E_w^l$ denote the number who enroll in the advanced course by the end of the simulation. Then $R_{bw}$ is defined as:
\begin{equation}
    R_{bw} = \frac{\frac{E_b^l}{N_b^l}}{\frac{E_w^l}{N_w^l}}
\end{equation}
An $R_{bw} = 1$ indicates that lower-class Black and White agents have equal likelihoods of enrolling in the advanced course. An $R_{bw} > 1$ indicates that lower-class Black agents are more likely to enroll than their White counterparts, whereas an $R_{bw} < 1$ indicates that they are less likely to do so.

\subsection{Computational procedures}

\subsubsection*{Advanced enrollment}

At each time step, each agent $i$ who is taken into consideration for advanced enrollment enrolls in the advanced course with probability $p_i$. If the agent is considered but does not enroll (which happens with probability $1 - p_i$), then the agent is no longer eligible for future consideration.

Thus, at each time step, advanced enrollment is governed by two procedures: (a) defining which agents are considered for enrollment, and (b) computing the enrollment probability $p_i$ for each considered agent.

Consideration depends on parameter $\beta_\textit{net-resource}$. If $\beta_\textit{net-resource} = 1$, then network-based resources matter for advanced enrollment. When that is the case, following Souto-Maior \citeyear{souto2025school}, the model defines that an agent is only considered for advanced enrollment if, and only if, $\textit{net-resource}_i = 1$, i.e., if they have access to the network-based resource. If $\beta_\textit{net-resource} = 0$, then network-based resources do not matter for advanced enrollment. In that case, all agents are considered for advanced enrollment.

Probability $p_i$ is defined as follows:
\begin{equation}
p_i = (1 - \beta_{\textit{admin-pressures}}) \; \textit{acad-background}_i + \beta_{\textit{admin-pressures}} \; \textit{influence-potential}_i
\label{eq:p_i}
\end{equation}

Let us unpack this equation. Following the course-taking literature reviewed above, this equation establishes that the agent's probability of advanced enrollment (given consideration) depends on a weighted average of their academic background, $\textit{acad-background}_i$, and their potential to shape gatekeepers' decisions, $\textit{influence-potential}_i$. Parameter $\beta_{\textit{admin-pressures}}$ captures the weight. Recall that this parameter captures the extent to which agents in this school engage in practices to pressure gatekeepers' decisions. This weight indicates that the more agents in this school engage in such practices, the more one's probability of advanced enrollment will depend on their potential to shape placement decisions.

For concreteness, consider the extreme cases of $\beta_{\textit{admin-pressure}}$. If agents in this school exert no pressure at all ($\beta_{\textit{admin-pressures}} = 0$), advanced enrollment decisions are determined entirely by academic background. Conversely, if pressure is maximized ($\beta_{\textit{admin-pressures}} = 1$), placement decisions depend solely on influence potential, meaning that enrollment outcomes are driven by agents' ability to sway gatekeepers. This formulation allows us to simulate contexts with varied intensity of agents' pressures, understanding how such exclusionary practices influence course-taking stratification.

Note that in this equation, whether engagement in administrative pressures influences one's probability of enrollment fully depends on one's potential to influence placement decisions (and vice-versa). This captures the notion that the extent to which parents (and students) can successfully advocate for education customization depends both on whether they actively pressure school staff and on how responsive gatekeepers are to their customization requests \cite{lewis2014inequality}.

Below, I detail the computational operationalization of the mechanisms of interest, explaining how I model agents' potential to shape placement decisions ($\textit{influence-potential}_i$) as well as how they gain access to network-based resources ($\beta_\textit{net-resource} = 0$).

\subsubsection*{Administrative pressures and the creation of White spaces}

As noted above, at high school entry, agents’ potential to influence course placement decisions is directly influenced by an individual-level attribute: social class. Because upper-class families often bring more financial resources to their districts and schools, gatekeepers have incentives to be more responsive to their customization requests \cite{cucchiara2013marketing, posey2014middle, lewis2014inequality}. Informed by this individual-level advantage, I defined (above) that, at high school entry, $\textit{influence-potential}_i = \textit{upper-class}_i$.

However, the literature on the creation of White spaces (reviewed above) illustrates that when agents pressure school personnel, they not only secure advantages for individual students but also influence the organizational context, fostering specific norms and expectations. When upper-class families are disproportionately White, as they pressure schools to customize educational settings, they---even if unintentionally---contribute to structuring schools around White norms and values. Then, when both Black and White families advocate for access to advanced coursework, they may face different chances of success.

To capture this process, once the model starts and students begin to navigate high school, I modify $\textit{influence-potential}_{i}$ so that it is endogenous to the high school context. At this stage, this variable is not solely defined by individual-level attributes but also by the interaction between structural conditions and the agent's race. Formally, I write:
\begin{equation}
\textit{influence-potential}_{i}  = 1 - 0.5 \; (1 -\textit{upper-class}_i) - 0.5\; \textit{black}_i \; \textit{class-ineq}_s
\label{eq:influence}
\end{equation}
Let us unpack this equation. The first term---the value of 1---captures the maximum value of $\textit{influence-potential}_{i}$. Then, other terms subtract from this maximum value. This subtraction captures the average of two effects.

The first subtraction term, the term $0.5 \; (1 -\textit{upper-class}_i)$, captures the individual-level effect defined above: the fact that social class directly conveys agents with a higher chance to influence gatekeepers' decisions. If the agent is from an upper-class background, then $\textit{upper-class}_i = 1$ and this subtraction term becomes 0. If the agent is from a lower-class background, then $\textit{upper-class}_i = 0$ and this subtraction term is maximized ($= 0.5$).

The second subtraction term, $\textit{black}_i \; \alpha_\textit{class-ineq}$, captures how agents’ engagement in administrative pressures can create White spaces. To understand the chosen operationalization, recall that the construction of a White space occurs when pressures to customize educational settings come from upper-class families that are disproportionately White. When this is the case, responses to pressures have the collateral effect of structuring schools around White norms and values. Therefore, whether Black agents have a lower likelihood of influencing placement decisions depends on the extent to which upper-class families are disproportionately White (parameter $\alpha_\textit{class-ineq}$). According to this framework, the equation above specifies that when there are no Black-White differences in social class, $\alpha_\textit{class-ineq} = 0$, the model does not initiate the creation of White spaces, and thus, this subtraction term becomes 0. As $\alpha_\textit{class-ineq}$ increases, the stronger the structuring of the school as a White space, decreasing Black agents' potential to shape gatekeepers' decisions.

\subsubsection*{Diffusion of network-based resources}

The model allows agents to gain access to the network-based resource ($\textit{net-resource}_i$) through social interactions with other agents in the school environment. Since each time step represents an opportunity for social interaction, the model defines that at every time step $> 0$, each agent $i$ with $\textit{net-resource}_i = 1$ randomly selects one of their ties (call it $j$) who does not already have access to the resource ($\textit{net-resource}_j = 0$) and shares the resource with a certain probability \cite{souto2025school}. As detailed below, this probability depends on the nature of the tie.

\subsubsection*{In-group favoritism}

The computational operationalization presented here expands on Souto-Maior's (\citeyear{souto2025school}) model to incorporate in-group favoritism. Specifically, if the race of $i$ and $j$ is the same, agent $i$ shares the resource with probability $\beta_\textit{diffusion-s}$. In contrast, if the race of $i$ and $j$ is different, agent $i$ shares the resource with probability $\beta_\textit{diffusion-d}$. Parameter $\beta_\textit{in-group-favoritism}$ captures the extent to which diffusion of resources varies across same- and different-race ties.

\section{Simulation results}

\subsection{The emergence of opportunity hoarding under no exclusionary behaviors}

The first goal of the model presented here is to demonstrate a limitation of the exclusionary-behaviors interpretation of opportunity hoarding: that exclusionary behaviors --- administrative influence and in-group favoritism --- are not necessary to generate race-based penalties in access to educational resources. The model shows how, through another relational process (diffusion of network-based resources), a race-based penalty in access to advanced coursework can arise even when agents do not engage in exclusionary behaviors.

To set up the simulation exercise, we first need to provide numerical values for the structural conditions that characterize the simulation environment. To define empirically reasonable values for social class parameters, recall that $\alpha_{\textit{class-ineq}} = 1 - \frac{\alpha_\textit{class-b}}{\alpha_\textit{class-w}}$. I initialize the model by defining values for $\alpha_{\textit{class-ineq}}$ and $\alpha_{\textit{class-w}}$, and computing $\alpha_\textit{class-b} = (1 - \alpha_\textit{class-ineq}) \, \alpha_\textit{class-w}$. For a given race group, the probability of being upper-class is defined as the probability that a randomly drawn agent from that race group is in the top \nth{10} percentile of the socioeconomic background composite reported by the Early Childhood Longitudinal Study, Kindergarten Cohort of 1998 (ECLS-K). Based on this definition,\footnote{\citet{fryer2004understanding} report that Black students have an average score of $-0.359$ and White students have an average score of $0.202$ in the 1998 ECLS-K composite of socioeconomic status. Given that these are standardized scores, it follows that a randomly drawn White student has about a 14\% chance of being in the top \nth{10} percentile, and a randomly drawn Black student has about a 5\% chance.} I set $\alpha_{\textit{class-w}} = 0.14$ --- that is, 14\% of White agents are from an upper-class background --- and $\alpha_{\textit{class-ineq}} = 0.74$, which implies $\alpha_\textit{class-b} = 0.05$ --- that is, 5\% of Black agents are from an upper-class background. Importantly, this initialization provides a reasonable baseline for the simulations. Results will later include sensitivity analyses for the level of racial disparities in social class background ($\alpha_{\textit{class-ineq}}$) considered in the simulated environment.

To define the level of racial homophily in the network, I follow \citet{currarini2010identifying}, who provide an empirical analysis of the National Longitudinal Study of Adolescent to Adult Health (Add Health). Based on their results,\footnote{Equation S2 of the Supporting Information file, computed for a relative group size of 50\%.} I set $\alpha_{\textit{racial-homophily}} = 0.5$. See \ref{A:net-model} for details on the construction of the network.

Given these structural conditions, the next step is to define the parameters that capture the mechanisms of interest. To simulate course-taking dynamics under different assumptions about agents’ engagement in exclusionary behaviors, I consider the following versions of the model:
\begin{itemize}
    \item \textbf{Administrative Pressures version}. This version simulates a context in which agents engage in administrative pressures. This occurs whenever $\beta_{\textit{admin-pressures}} > 0$. To illustrate one of these possibilities, I set $\beta_{\textit{admin-pressures}} = 0.5$. Further, this version isolates administrative pressures as the sole micro-level process influencing course-taking dynamics, assuming that network-based resources do not shape placement decisions ($\beta_{\textit{net-resource}} = 0$) and that agents do not display in-group favoritism ($\beta_{\textit{in-group-favoritism}} = 0$).
    \item \textbf{In-group Favoritism version}. This version simulates a context in which agents display some level of in-group favoritism in their social interactions. This occurs whenever $\beta_{\textit{in-group-favoritism}} > 0$. To illustrate one of these possibilities, I set $\beta_{\textit{in-group-favoritism}} = 0.5$.\footnote{Further, to initialize the diffusion process, I define $\beta_\textit{diffusion-s} = 0.1$, meaning that agents have a 10\% chance of successfully sharing the resource with the chosen tie. Then, given $\beta_\textit{diffusion-s} = 0.1$; $\beta_\textit{in-group-favoritism} = 0.5$, and since $\beta_\textit{diffusion-d} = (1 - \beta_\textit{in-group-favoritism}) \, \beta_\textit{diffusion-s}$, it follows that $\beta_\textit{diffusion-d} = 0.05$} Because the role of in-group favoritism presumes that network-based resources matter for advanced enrollment, I set $\beta_{\textit{net-resource}} = 1$. Importantly, this version assumes that agents do not engage in administrative pressures ($\beta_{\textit{admin-pressures}} = 0$).
    \item \textbf{No Exclusionary Behaviors version}. This version simulates how the diffusion of network-based resources ($\beta_{\textit{net-resource}} = 1$) shapes course-taking dynamics even when agents do not engage in any exclusionary behaviors ($\beta_{\textit{admin-pressures}} = 0$ and $\beta_{\textit{in-group-favoritism}} = 0$).
\end{itemize}

Table~\ref{tab:combinations1} summarizes the parameter combinations for each of the model versions of interest.

\begin{table}[h!]
    \centering
    \caption{Parameter combinations for model versions of interest}
    \label{tab:combinations1}
    \begin{tabular}{l c c c}
    \hline
    \addlinespace[0.3em]
     & \textbf{Administrative} & \textbf{In-group} & \textbf{No Exclusionary} \\
    & \textbf{Pressures} & \textbf{Favoritism} & \textbf{Behaviors} \\
    \addlinespace[0.3em]
    \hline
    \addlinespace[0.5em]
    \multicolumn{4}{l}{\textbf{Structural conditions}} \\
    \addlinespace[0.5em]
    $\alpha_{\textit{class-ineq}}$ & 0.74 & 0.74 & 0.74 \\ 
    $\alpha_{\textit{racial-homophily}}$ & 0.5 & 0.5 & 0.5 \\
    \addlinespace[1em]
    \multicolumn{4}{l}{\textbf{Micro-level processes}} \\
    \addlinespace[0.5em]
    $\beta_{\textit{admin-pressures}}$ & 0.5 & 0 & 0 \\ 
    $\beta_{\textit{net-resource}}$ & 0 & 1 & 1 \\ 
    $\beta_{\textit{in-group-favoritism}}$ & 0 & 0.5 & 0 \\ 
    \addlinespace[0.3em]
    \hline
    \addlinespace[0.3em]
    \multicolumn{4}{l}{\footnotesize \textit{Note:} Additional parameters used in the initialization of the model are: $\alpha_\textit{class-w} = 0.14$} \\
    \multicolumn{4}{l}{\footnotesize and $\beta_\textit{diffusion-s} = 0.1$.} \\
    \end{tabular}
\end{table}
\FloatBarrier

\begin{figure}[h!]
    \centering
    \includegraphics{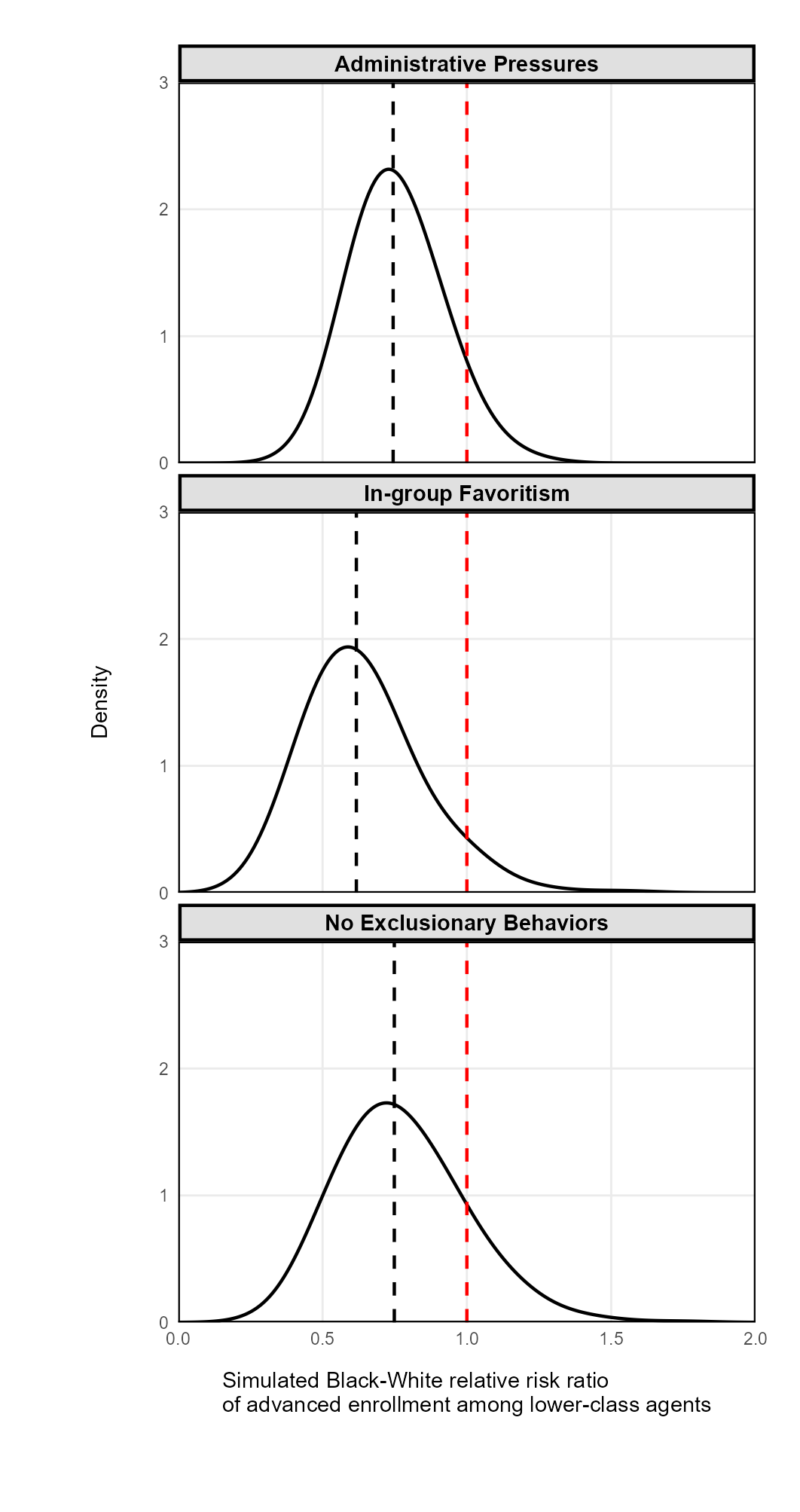}
    \caption{\textbf{Distribution of simulated race-based penalties under the Administrative Pressures, In-group Favoritism, and No Exclusionary Behaviors versions of the model.} Parameters for each of these versions are defined in Table~\ref{tab:combinations1}. Each version is simulated 1,000 times. Solid lines report the distribution of the Black-White relative risk ratio of advanced enrollment for lower-class agents, $R_{bw}$. Vertical dotted lines capture the average value of $R_{bw}$ across the 1,000 simulation runs. The vertical red line indicates $R_{bw} = 1$ for reference.}
    \label{fig_micro-processes}
\end{figure}

The final step in this exercise is to simulate course-taking dynamics in each of these model versions of interest. Accounting for the stochasticity of the agent-based model, I simulate each combination of interest 1,000 times. Figure~\ref{fig_micro-processes} presents the results.

\FloatBarrier

Results from the Administrative Pressures and In-group Favoritism versions of the model show that agents’ engagement in exclusionary behaviors --- either through administrative pressures or in-group favoritism --- tends, under empirically reasonable structural conditions, to produce a race-based penalty in advanced enrollment. These results indicate that the model developed here captures the insight emphasized by the exclusionary-behaviors interpretation of opportunity hoarding: exclusionary behaviors are a central mechanism through which White actors can hoard educational resources.

However, in contrast to the exclusionary-behaviors interpretation, results from the No Exclusionary Behaviors version of the model show that race-based penalties in access to advanced coursework emerge even when agents are simulated not to display exclusionary behaviors.

To better understand this notion of hoarding without hoarders, let us unpack how the computational model generates this result. Recall that, in the model designed here, agents’ chances of advanced enrollment are a function of two procedures: (a) determination of which agents are considered for advanced enrollment, and (b) probability of enrollment given consideration (Equation~\ref{eq:p_i}). In the No Exclusionary Behaviors version of the model, $\beta_\textit{admin-pressures} = 0$ and, thus, following Equation~\ref{eq:p_i}, $p_i = \textit{acad-background}_i$. Because, by design, lower-class Black and White agents enter high school with the same levels of academic preparation, it follows that lower-class Black and White agents have the same probability of enrollment given consideration. Therefore, the formation of race-based penalties stems only from Black-White differences generated through procedure (a) --- determination of which agents are considered for advanced enrollment.

Such consideration depends on whether the agent has access to the network-based resource needed for advanced enrollment. To understand how lower-class Black-White differences in access to this resource emerge, recall that the model does not presuppose racial penalties from the start --- at time step 0, no lower-class (neither Black nor White) agent has access to the network-based resource, and all upper-class agents (Black or White) are endowed with access to this resource. Differences in access to the network-based resource between lower-class Black and White agents are thus the product of the extent to which agents are able to gain access to the resource through social interactions with others in the community.

\begin{figure}[h!]
    \centering
    \includegraphics{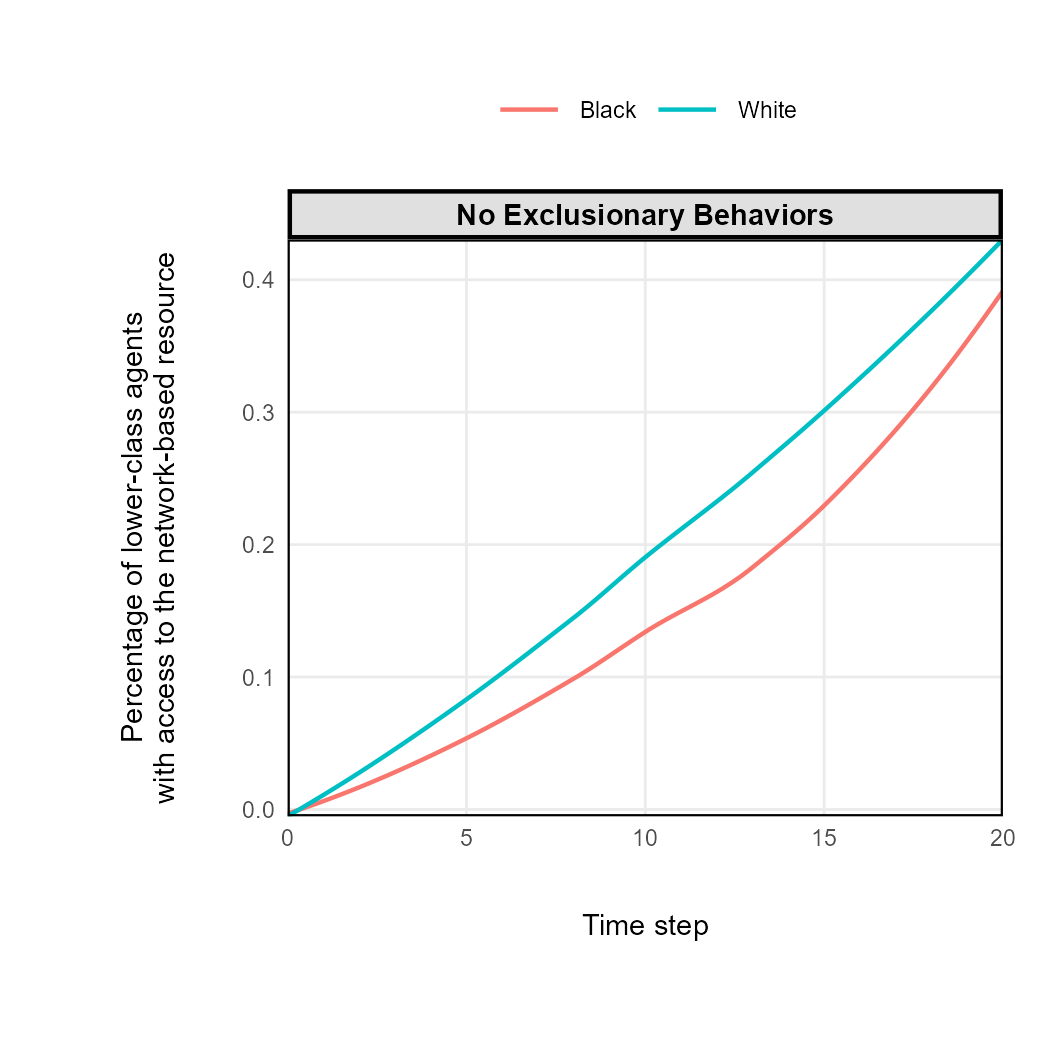}
    \caption{\textbf{Resource diffusion across time steps for the No Exclusionary Behaviors version of the model.} The figure plots smoothed trend curves for the share of lower-class agents (by race) with access to the network-based resource across the model’s time steps. The curves are smoothed over 1,000 simulation runs of the No Exclusionary Behaviors version defined in Table~\ref{tab:combinations1}.}
    \label{fig_time-steps}
\end{figure}

Given these initial conditions, lower-class agents need to interact with upper-class agents to gain access to the network-based resource. Capturing empirically representative structural conditions, White agents are more likely to be from an upper-class background than Black agents. This structural condition, combined with the fact that agents' networks are racially segregated ($\beta_\textit{racial-homophily} = 0.5$), implies that lower-class White agents have access to a more resourceful pool of network ties than Black agents. Then, through the social diffusion of network-based resources, lower-class White agents tend to gain faster access to the network-based resource than lower-class Black agents, securing enrollment in the advanced course before Black agents. Figure~\ref{fig_time-steps}, below, helps us visualize this dynamic.

\FloatBarrier

Figure~\ref{fig_time-steps} captures lower-class Black and White agents' access to the network-based resource across simulated time steps. At time step 0, by design, no lower-class Black or White agent has access to this resource. However, once the simulations begin, agents are allowed to share resources with their ties, and lower-class agents gradually gain access to the resource through interactions with upper-class agents. Because lower-class White agents have access to a more resourceful pool of network ties than lower-class Black agents, White agents tend, at least initially, to gain access to the resource at a faster rate --- note that up until around time step 12, the trend curve for White agents is steeper. Early access to the resource is essential, as the number of available advanced course seats is limited and quickly fills up. Such early advantage, therefore, generates a higher likelihood of advanced enrollment for lower-class White agents. After approximately time step 12, the slope of the trend curve for Whites and Blacks become similar and, in fact, after this point, Black agents’ access to the network-based resource increases at a faster rate.\footnote{To understand this pattern, note that not all agents have the same number of ties. Number of ties is uniformly distributed for both Black and White agents (see the network formation model in \ref{A:net-model}). Then, this pattern indicates that around time step 12, most well-connected (or easy-to-reach) White agents already gained access to the resource. However, there are many well-connected Black agents who do not yet have access to it. As a result, while opportunities for resource reception for White agents begin to slow down, opportunities for resource reception among Black agents are still fast.} However, when this occurs, most spots in the model are already taken, and White agents have already secured an advantage in advanced enrollment.

\subsection{Unpacking the conditions under which such micro-level processes can produce race-based penalties}

The second goal of the model is to clearly articulate the conditions under which the micro-level processes of interest generate racial penalties in advanced enrollment (what I defined earlier as task (3)). Micro-level processes are context-dependent and can generate different outcomes depending on the structural context in which they are embedded. Therefore, this section provides a simulation exercise to assess how variations in the model's structural conditions --- racial differences in social class and racial homophily in agents' networks --- shape model outcomes.

\begin{figure}[h!]
    \centering
    \includegraphics{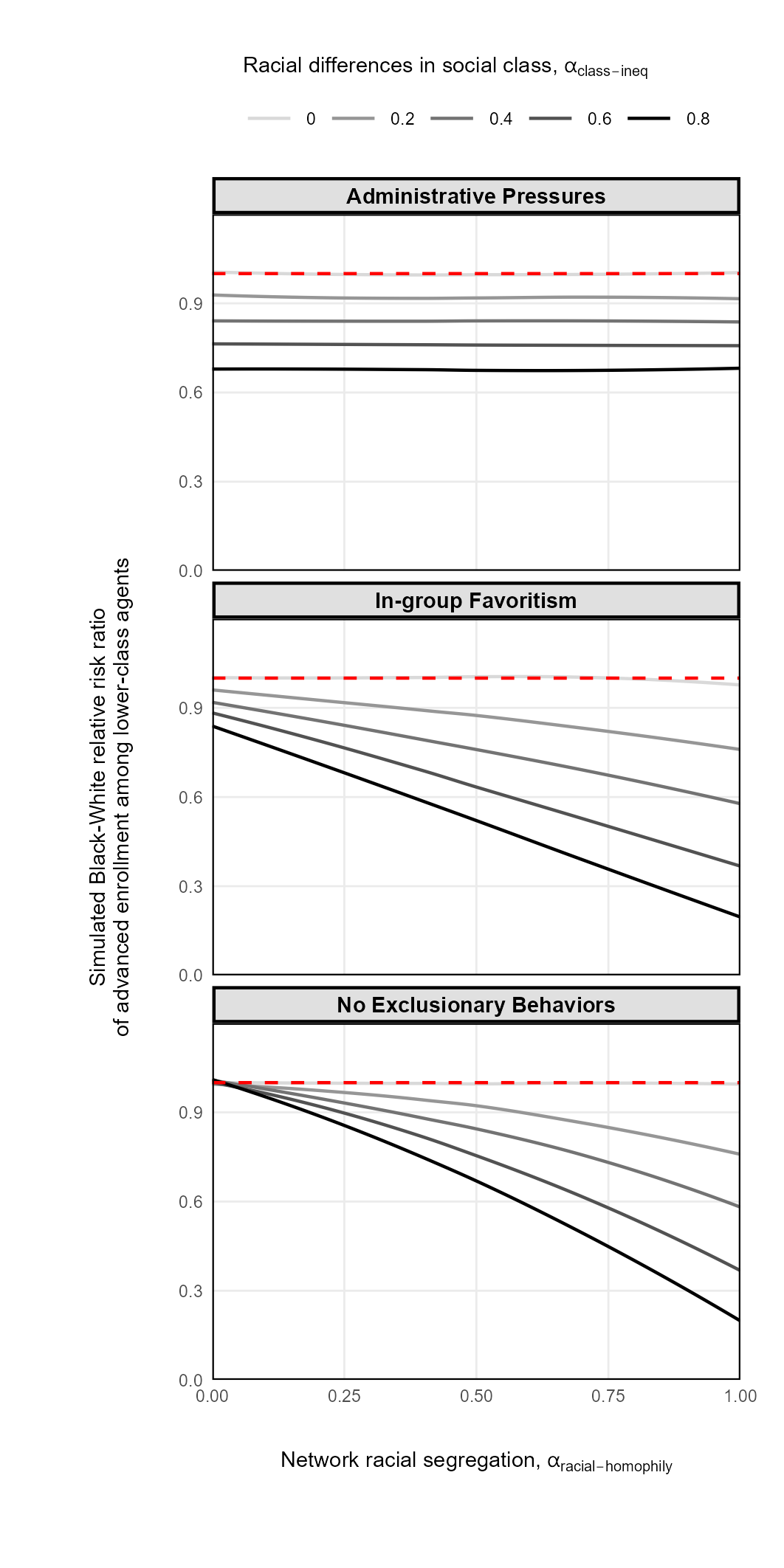}
    \caption{\textbf{Simulated race-based penalties from the Administrative Pressures, In-group Favoritism, and No Exclusionary Behaviors versions of the model under variations in structural conditions --- racial differences in social class and network racial segregation.} Parameters for each of these versions are defined in Table~\ref{tab:combinations-2}. Each parameter combination is simulated 1,000 times. Trend curves are smoothed across all simulated outcomes from each parameter combination. The horizontal red line indicates $R_{bw} = 1$  for reference.}
    \label{fig_contextual-conditions}
\end{figure}

Here, I consider the same model versions defined above --- Administrative Pressures, In-group Favoritism, and No Exclusionary Behaviors --- but allow structural parameters $\alpha_{\textit{class-ineq}}$ and $\alpha_{\textit{racial-homophily}}$ to vary. Table~\ref{tab:combinations-2} summarizes the new parameter combinations of interest and Figure~\ref{fig_contextual-conditions} presents the results for 1,000 simulations of each of these combinations.
    
\begin{table}[h!]
    \centering
    \caption{Parameter combinations for model versions of interest: sensitivity to structural conditions}
    \label{tab:combinations-2}
    \begin{tabular}{l c c c}
    \hline
    \addlinespace[0.3em]
     & \textbf{Administrative} & \textbf{In-group} & \textbf{No Exclusionary} \\
    & \textbf{Pressures} & \textbf{Favoritism} & \textbf{Behaviors} \\
    \addlinespace[0.3em]
    \hline
    \addlinespace[0.5em]
    \multicolumn{4}{l}{\textbf{Structural conditions}} \\
    \addlinespace[0.5em]
    $\alpha_{\textit{class-ineq}}$ & $\{0, 0.2, \ldots, 0.6, 0.8\}$ & $\{0, 0.2, \ldots, 0.6, 0.8\}$ & $\{0, 0.2, \ldots, 0.6, 0.8\}$ \\ 
    $\alpha_{\textit{racial-homophily}}$ & $\{0, 0.1, \ldots, 0.9, 1\}$ & $\{0, 0.1, \ldots, 0.9, 1\}$ & $\{0, 0.1, \ldots, 0.9, 1\}$ \\
    \addlinespace[1em]
    \multicolumn{4}{l}{\textbf{Micro-level processes}} \\
    \addlinespace[0.5em]
    $\beta_{\textit{admin-pressures}}$ & 0.5 & 0 & 0 \\ 
    $\beta_{\textit{net-resource}}$ & 0 & 1 & 1 \\ 
    $\beta_{\textit{in-group-favoritism}}$ & 0 & 0.5 & 0 \\ 
    \addlinespace[0.3em]
    \hline
    \addlinespace[0.3em]
    \multicolumn{4}{l}{\footnotesize \textit{Note:} Additional parameters used in the initialization of the model are: $\alpha_\textit{class-w} = 0.14$} \\
    \multicolumn{4}{l}{\footnotesize and $\beta_\textit{diffusion-s} = 0.1$.} \\
    \end{tabular}
\end{table}

\FloatBarrier
This exercise details how the effects of the micro-level processes of interest on the production of race-based penalties in advanced enrollment are context-dependent. This helps us more clearly articulate the dynamics of these mechanisms. Let us unpack results for each model version.

In the Administrative Pressures version of the model, we observe that race-based penalties arise only when racial differences in social class are present (the lighter gray lines, where $\alpha_{\textit{class-ineq}} = 0$, are flat around $R_{bw} = 1$) and that network racial segregation does not shape model dynamics (all lines have slope 0). With respect to racial differences in social class, this result is consistent with theoretical accounts of the creation of White spaces, which posit that White spaces are constructed as disproportionately White upper-class families advocate for educational customization. Computationally, this pattern stems from the formulation of $\textit{influence-potential}_i$ (Equation~\ref{eq:influence}), which defines racial differences in agents' potential to shape gatekeepers' decisions as increasing linearly with $\alpha_{\textit{class-ineq}}$. With respect to network racial segregation, this is also consistent with the theoretical framework, as the creation of White spaces depends on agents' direct interactions with gatekeepers and is unaffected by the structure of agents' peer networks.

The In-group Favoritism and No Exclusionary Behaviors versions of the model display similar dynamics. In both versions, network-based resources influence advanced enrollment, and agents do not engage in administrative pressures ($\beta_{\textit{admin-pressures}} = 0$). As a result, these model variants exhibit the computational mechanisms discussed in the prior section: simulated race-based penalties arise solely from differences in Black and White agents' interactions with upper-class agents. In the No Exclusionary Behaviors version, because agents do not display in-group favoritism, race-based penalties depend exclusively on differences in expected contact with upper-class agents. These differences emerge from the joint effects of racial disparities in social class ($\alpha_{\textit{class-ineq}} > 0$) and racial segregation within social networks ($\alpha_{\textit{racial-homophily}} > 0$). Accordingly, when either structural condition is absent ($\alpha_{\textit{class-ineq}} = 0$ or $\alpha_{\textit{racial-homophily}} = 0$), no race-based penalties emerge ($R_{bw} = 1$).

In the In-group Favoritism version, a similar dynamic occurs, but in this case, simulated race-based penalties depend both on Black and White agents' expected contact with upper-class agents and on the tendency of agents to favor same-race peers in resource diffusion. The in-group favoritism mechanism amplifies the inequalities generated by differences in contact with upper-class agents: even when contact with upper-class agents exists, Black agents are less likely to gain access to the resource if upper-class agents (who are disproportionately White) prioritize supporting same-race peers. As a result, race-based penalties are larger and more persistent than in the No Exclusionary Behaviors version, especially when racial segregation in networks ($\alpha_{\textit{racial-homophily}}$) is low. In fact, note that this dynamic allows the in-group favoritism mechanism to generate race-based penalties even when there is no network racial segregation ($\alpha_{\textit{racial-homophily}} = 0$). It follows, therefore, that while both structural conditions are necessary for the No Exclusionary Behaviors version to produce race-based penalties, only racial differences in social class are needed for the In-group Favoritism version to do so.

Taken together, these results highlight how the different relational mechanisms of interest depend on structural conditions to generate race-based penalties. Exclusionary behaviors --- both the administrative influence and in-group favoritism mechanisms --- only require racial differences in social class to produce some race-based penalty. In contrast, the network diffusion process underlying hoarding without hoarders requires the byproduct of racial differences in social class and network racial segregation to do so. 

\section{Conclusion}

This theoretical paper set out to clarify the concept of opportunity hoarding, particularly as it applies to the formation of Black-White educational inequalities. Despite its widespread use, opportunity hoarding is often loosely defined, leading to varied interpretations of the inequality-producing mechanisms it captures. I identified two prevailing interpretations. First, the group-disparity interpretation, which defines opportunity hoarding in terms of observed outcomes, allowing any micro-level process that generates group disparities to fall under the umbrella of opportunity hoarding. Second, the exclusionary-behaviors interpretation, which defines opportunity hoarding in terms of the ways in which White individuals engage in exclusionary practices --- captured by the mechanisms of in-group favoritism and administrative pressures --- to secure access to valuable resources. I have argued that both interpretations fail to fully capture the conceptual potential of opportunity hoarding.

Building on Tilly's \citeyearpar{tilly1998durable} original articulation, this paper proposed a more precise interpretation: racial opportunity hoarding captures the relational processes that generate racial disparities in access to valuable resources \textit{that cannot be explained by differences in individual attributes} --- that is, disparities which arise between different-race individuals of similar characteristics (race-based penalties). Importantly, while opportunity hoarding can refer to both the \textit{emergence} and the \textit{reproduction} of group-based penalties, for conceptual precision, the focus here was on the \textit{emergence and not reproduction}. Future formulations can explore the distinct relational processes underlying the reproduction of existing group-based penalties.

From this definition, the paper argued that the group-disparity interpretation is too broad, as it fails to emphasize how opportunity hoarding highlights relational micro-processes that contrast with individualist explanations for group disparities. Conversely, the exclusionary-behaviors interpretation is too narrow, overlooking the fact that exclusionary behaviors are not the only relational mechanisms that can produce race-based penalties. Another relational process --- the diffusion of network-based resources --- can, when combined with racial differences in social class and network segregation, generate race-based penalties even when behaviors are race-neutral, a dynamic I conceptualize as \textit{hoarding without hoarders}.

To clearly demonstrate the proposed interpretation of opportunity hoarding as well as the notion of hoarding without hoarders, the paper constructed an agent-based model of opportunity hoarding in education. For concreteness, the model focused on one context: racial disparities in access to advanced high school coursework between students in the same school. By formalizing the key micro-level processes of interest, the model showed how, through the diffusion of network-based resources, race-based penalties in advanced enrollment can emerge even in the absence of administrative pressures or in-group favoritism. Further, by unpacking the model dynamics, it showed how hoarding without hoarders emerges as a byproduct of racial differences in social class and network racial segregation. 

In reflecting on this contribution, it is important to recognize its limitations, which also point toward directions for future research. For instance, the model constructed here is highly stylized---e.g., it considers only two race groups, a heterogeneous school racial composition, and a fixed network structure. These idealizations were introduced to minimize unnecessary complexities, helping to provide a clear illustration of the concept of hoarding without hoarders. However, future research can build on the modeling approach presented here by examining, empirically or theoretically, the conditions under which hoarding without hoarders is more likely to arise. In particular, it would be valuable to explore variations in network structure, as the structure of agents' networks is not exogenous to educational contexts: school organization can influence the frequency, duration, formality, and cooperative nature of social interactions \cite{small2009unanticipated}. Such extensions could provide insights into how educational organizations might counteract the emergence hoarding without hoarders. Moreover, while this paper focused on advanced course-taking as one instance of opportunity hoarding, future research should examine the hoarding of other educational resources, such as access to well-resourced schools or districts. Finally, the general framework proposed here could also be expanded beyond education to other domains where opportunity hoarding plays a central role, including employment and housing.

Overall, this paper contributes to a more consistent and systematic conceptualization of opportunity hoarding as an explanation for educational disparities. In particular, by proposing the notion of hoarding without hoarders, it suggests that the common emphasis on exclusionary behaviors understates the broader structural conditions under which White actors can hoard valuable educational resources. This highlights that, as scholars and policymakers seek to document, measure, and design efforts to counteract opportunity hoarding, they must not be constrained by a narrow focus on identifying exclusionary behaviors alone. Instead, they must also consider the structural interplay of network segregation and racialized inequality in the hoarding of educational opportunities across racial lines.




\FloatBarrier
\pagebreak
\begin{singlespacing}
\bibliography{references-school-effects.bib,
              references-course-taking.bib,
              references-peer-effects.bib,
              references-discrimination.bib, 
              references-ABM.bib, 
              references-opportunity-hoarding.bib}
\bibliographystyle{apalike}
\end{singlespacing}
\renewcommand{\bibname}{References}



\FloatBarrier
\pagebreak

\renewcommand{\thesection}{Appendix A}
\section{}
\label{A:net-model}
\renewcommand{\thesection}{A}
\renewcommand\thefigure{\thesection.\arabic{figure}} 
\renewcommand\thetable{\thesection.\arabic{table}} 
\setcounter{figure}{0}  
\setcounter{table}{0}  

\section{Formation of agents' networks}

The purpose of the network formation procedure is to endow agents with racially segregated networks, where network segregation is defined by Coleman's \citeyearpar{coleman1958relational} inbreeding homophily index, $IH_g$. Given a group $g$ (e.g., a racial group), the relative group size of group $g$ in the school ($S_g$), and the likelihood that agents of group $g$ form same-group ties ($Q_g$), Coleman’s inbreeding homophily index for group $g$ is defined as:
\begin{equation}
IH_g = \frac{Q_g - S_g}{1 - S_g}.
\label{eq:IH_def}
\end{equation}

To endow agents with racially segregated networks, the model presented here, following \citeauthor{souto2025school}, uses the following procedure: each agent $i$ from group $g$ establishes $n_i$ undirected ties with other agents in the environment, ensuring that a proportion $\gamma_g$ of these ties consist of same-race connections. Each agent’s number of ties, $n_i$, is drawn from a discrete uniform distribution on $\{0,1,\ldots,\theta\}$. To ensure an empirically reasonable average number of ties, I set $\theta = 8$. This value is chosen based on an empirical calibration procedure that selected the $\theta$ producing an average number of ties per agent closest to 6.7 --- following \citeauthor{currarini2010identifying}'s (2010) analysis of the National Longitudinal Study of Adolescent Health, this is the estimated number of undirected friendship ties in U.S. high schools when relative group size is 50\%.

This network-formation procedure ensures that the simulated average share of same-group ties equals $\gamma_g$. To initialize the model at a desired level of racial homophily ($IH_g$), we can define $\gamma_g$ by solving Equation~\ref{eq:IH_def} for $Q_g$ and replace $Q_g$ with $\gamma_g$:
\begin{equation}
\gamma_g = IH_g (1 - S_g) + S_g.
\label{eq:gamma_from_IH}
\end{equation}

Given that, in the model designed here, each racial group has the same relative size (50\%), we can replace $S_g$ with 0.5. Further, we can use a single inbreeding homophily parameter ($\alpha_\textit{racial-homophily}$) for each group. Then, we define $\gamma$ based on $\alpha_\textit{racial-homophily}$ by writing:
\begin{align}
\gamma &= \alpha_\textit{racial-homophily} (1 - 0.5) + 0.5.
\label{eq:gamma_equal_groups}
\end{align}

\begin{figure}[h!]
	\centering
	\includegraphics{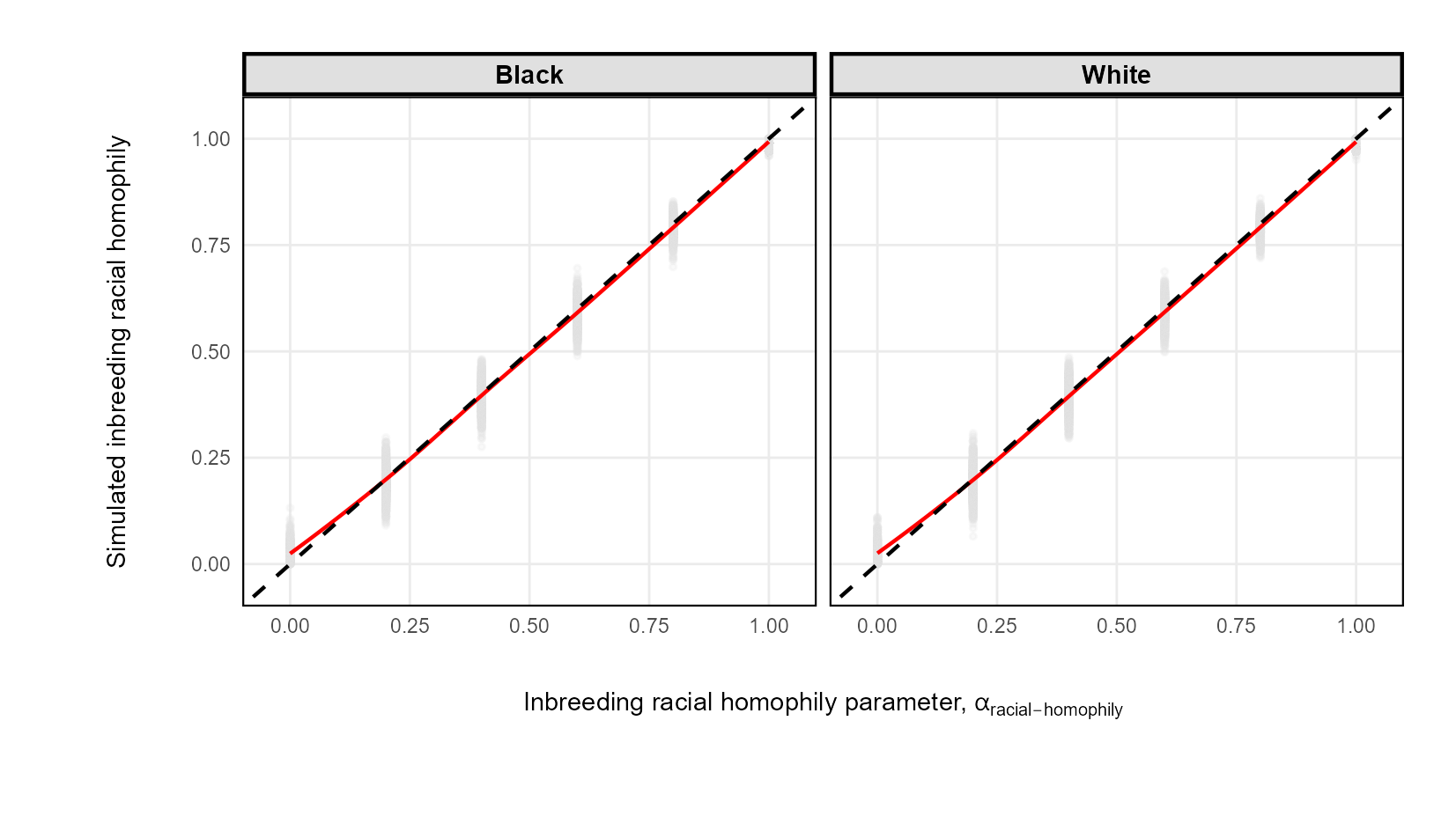}
	\caption{\textbf{Target and simulated inbreeding racial homophily for Black and White agents.} Results from 1,000 simulation runs for each value of $\alpha_\textit{racial-homophily} \in \{0, 0.2, \ldots, 0.8, 1\}$. Red trend curves are smoothed across all simulated outcomes from each parameter combination. Number of agents follows the model in the main text ($= 400$); $\theta = 8$.}
	\label{fig_net-model-IH}
\end{figure}

Figure~\ref{fig_net-model-IH} demonstrates how this formulation allows the parameter $\alpha_\textit{racial-homophily}$ to produce the desired level of inbreeding racial homophily for both groups.

\begin{figure}[h!]
	\centering
	\includegraphics{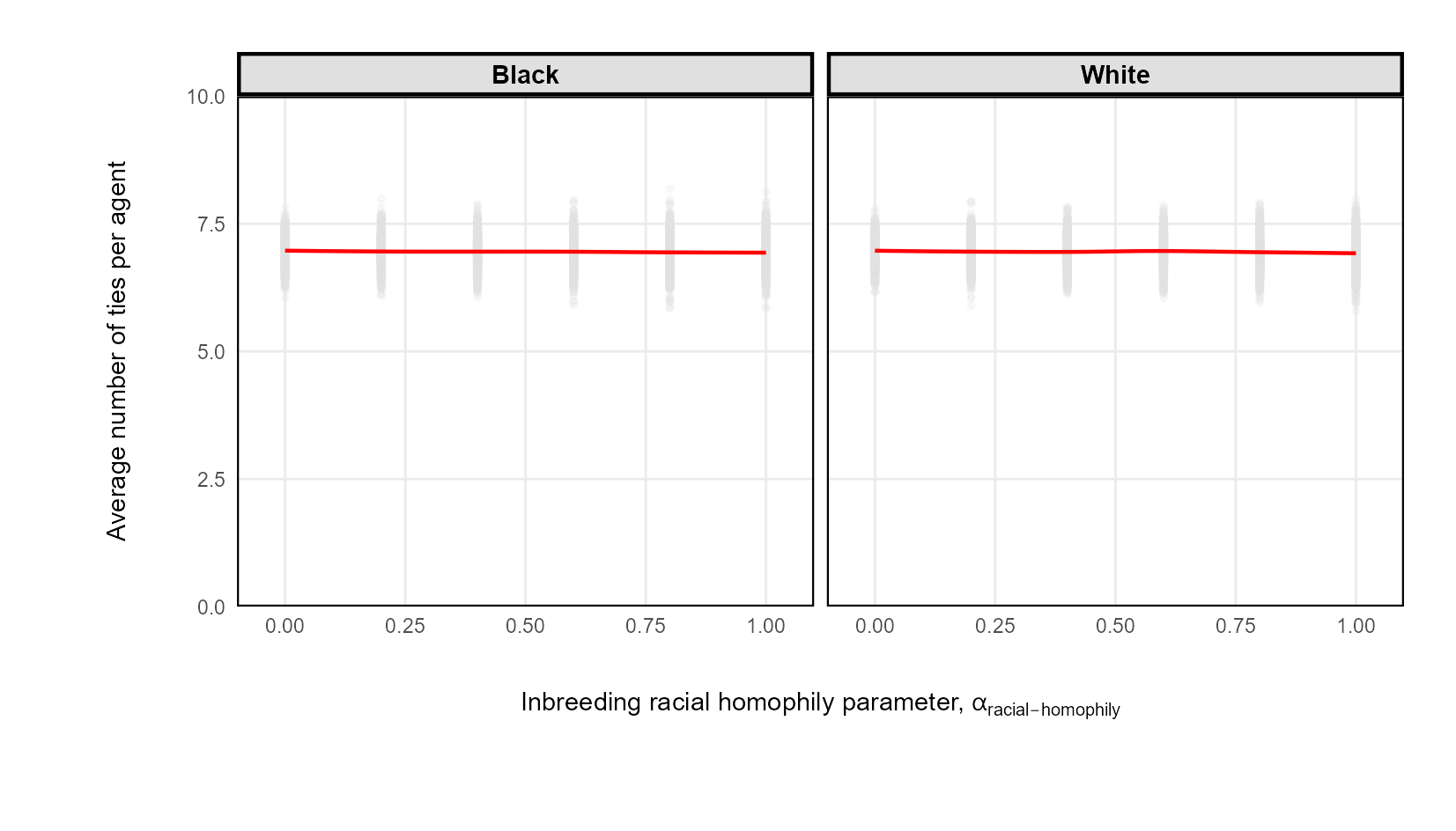}
	\caption{\textbf{Simulated average number of undirected ties per agent, by racial group.} Results from 1,000 simulation runs for each value of $\alpha_\textit{racial-homophily} \in \{0, 0.2, \ldots, 0.8, 1\}$. Red trend curves are smoothed across all simulated outcomes from each parameter combination. Number of agents follows the model in the main text ($= 400$); $\theta = 8$.}
	\label{fig_net-model-IH-n-ties}
\end{figure}

Further, Figure~\ref{fig_net-model-IH-n-ties} shows that the average simulated number of undirected ties per agent closely approximates the target value of 6.7; that it is equally distributed across racial groups; and that it remains constant across values of the inbreeding racial homophily parameter, $\alpha_\textit{racial-homophily}$.

\end{document}